\documentclass{article}
\usepackage{spconf,amsmath,graphicx}
\usepackage{pifont}
\usepackage{bm}
\usepackage{amsfonts,amssymb}
\usepackage{algorithm}
\usepackage{algorithmic}

\title{Emotion-Aware Contrastive Adaptation Network for Source-free Cross-corpus Speech Emotion Recognition}
\name{Yan Zhao$^{1,2,\#}$\thanks{ $^{*}$ Corresponding authors, $^{\#}$ Contribute equally to this work.} \thanks{This work was supported in part by the National Key R \& D Project under the Grant 2022YFC2405600, in part by the NSFC under the Grant U2003207 and 61921004, in part by the Jiangsu Frontier Technology Basic Research Project under the Grant BK20192004, in part by the YESS Program by CAST under the Grant 2023QNRC001, and in part by the ASFC under the Grant 2023Z071069003.}, Jincen Wang$^{2,3,\#}$, Cheng Lu$^{2,3}$, Sunan Li$^{1,2}$, Bj$\ddot{o}$rn W. Schuller$^{4,5}$, \textit{Yuan Zong}$^{2,3,*}$, \textit{Wenming Zheng}$^{2,3,*}$}
\address{$^1$School of Information Science and Engineering, Southeast University, Nanjing, China\\
$^2$Key Laboratory of Child Development and Learning Science of Ministry of Education, \\Southeast University, Nanjing, China\\
$^3$School of Biological Science and Medical Engineering, Southeast University, Nanjing, China\\
$^4$GLAM – Group on Language, Audio, \& Music, Imperial College London, UK \\
$^5$Chair of Embedded Intelligence for Health Care
and Wellbeing, University of Augsburg, Germany\\
\{xhzongyuan, wenming\_zheng\}@seu.edu.cn
}

\begin{document}
\ninept
\maketitle
\begin{abstract}
Cross-corpus speech emotion recognition (SER) aims to transfer emotional knowledge from a labeled source corpus to an unlabeled corpus. However, prior methods require access to source data during adaptation, which is unattainable in real-life scenarios due to data privacy protection concerns.  This paper tackles a more practical task, namely source-free cross-corpus SER, where a pre-trained source model is adapted to the target domain without access to source data. To address the problem, we propose a novel  method called  emotion-aware contrastive adaptation network (ECAN). The core idea is to capture local neighborhood information between samples while considering the global class-level adaptation. Specifically,  we propose a nearest neighbor contrastive learning to promote local emotion consistency among features of highly similar samples. Furthermore, relying solely on nearest neighborhoods may lead to ambiguous boundaries between clusters. Thus, we incorporate supervised contrastive learning to encourage greater separation between clusters representing different emotions, thereby facilitating improved class-level adaptation. Extensive experiments indicate that our proposed ECAN significantly outperforms state-of-the-art methods under the source-free cross-corpus SER setting on several speech emotion corpora.

\end{abstract}
\begin{keywords}
Source-free cross-corpus speech emotion recognition, speech emotion recognition, contrastive learning, transfer learning.
\end{keywords}
\section{Introduction}
\label{sec:intro}

Over the course of the last decade, 
diverse SER applications have gained lots of attention with  tremendous progress of deep learning \cite{wagner2023dawn, lu2022domain, akccay2020speech, lu2022speech}. Though achieving huge successes, conventional SER methods may encounter performance degradation, even when the training data and the test data deviate slightly from each other. Thus, researchers turn their attention to cross-corpus SER, where the training data and test data come from different corpora, and multiple methods have been proposed for cross-corpus SER \cite{ahn2021cross, 10095388}. Conventional cross-corpus SER methods trend to alleviate the domain discrepancy by domain matrices \cite{zhao2022deep, zhang2021cross} or adversarial training \cite{9756868, 9695273}. Taking the adversarial training based methods as an example, this kind of methods obfuscates the domain discriminator to prevent it from distinguishing between the source and target corpus samples.

Common cross-corpus SER algorithms assume all data is available during adaptation. In real-life scenarios, this assumption is rarely possible due to data privacy protection. Labeled emotional voices can be treated as a form of identification for specific individuals \cite{nagrani2018seeing}. Improper disclosure of data with corresponding labels could unduly influence data providers. This paper focuses on a more practical and interesting task called source-free cross-corpus SER, where source data is inaccessible during adaptation. The goal is to adapt a pre-trained model, originally trained on a source corpus, to perform well on a target corpus without any labeled source data. Traditional cross-corpus SER methods focus on matching the target feature distribution with the source one to alleviate domain gaps. However, 
here, 
this does not work since source-free cross-corpus SER faces the challenge of source distribution estimation without access to the source data.  In this context, the main dilemma of this task is how to effectively utilize the pre-trained source model to identify target samples correctly despite the presence of domain shifts. 

\begin{figure*}[t!] 
\centering 
\includegraphics[width=0.88\textwidth]{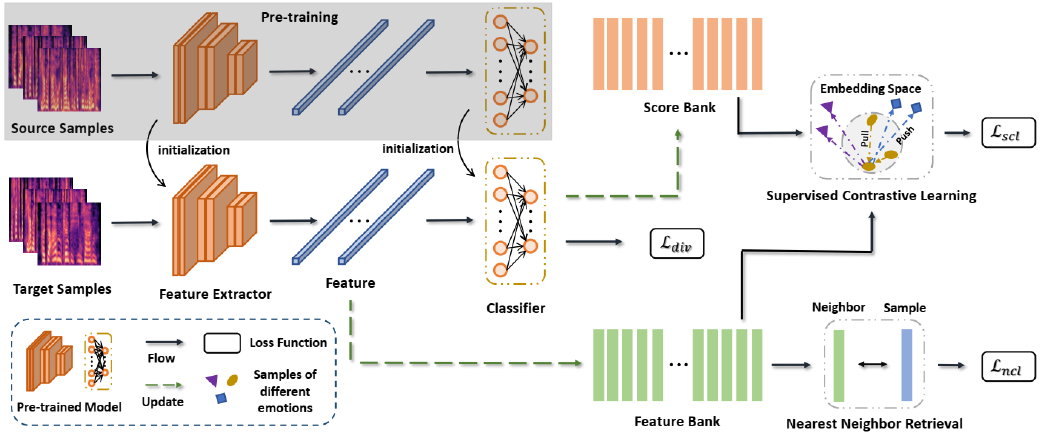}\\ 
\caption{Overview Structure of the Proposed ECAN in Dealing with Source-free Cross-Corpus SER.}
\label{fig1} 
\end{figure*}

To tackle the more practical and previously unexplored source-free cross-corpus SER problem, we propose a simple, yet effective method called emotion-aware contrastive adaptation network (ECAN). It is worth noting that this is the first work dedicated to addressing the source-free cross-corpus SER problem. The key idea is to update the target model using the pre-trained source model from both local and global perspectives. Building upon previous research \cite{liang2020we, yang2021exploiting}, we find that target data of the same emotion tends to form a cluster in the feature space despite domain shifts between source and target corpora. To exploit this inherent local structure between target data, we propose a novel nearest neighbor contrastive learning algorithm to enhance the semantic consistency among neighboring samples. Besides, solely relying on nearest neighbor information for adaptation may result in ambiguous category boundaries, as the local structure may not capture the full complexity of emotion distribution within the target data. To address this limitation, we incorporate supervised contrastive learning into the network, which aims to pull away clusters of different emotions, to achieve emotion-wise global adaptation. Both nearest neighbor and supervised contrastive learning modules work in a 
promotion way, jointly considering enhancing of nearest neighbor information and emotion-level global adaptation.


\section{PROPOSED METHOD}
\label{sec:format}

In this section, we will present the details of the proposed ECAN for coping with the problem of source-free cross-corpus SER. Formally, we denote the labeled source speech emotion corpus and unlabeled target one as ${{\mathcal{D}}_{s}}$ and ${{\mathcal{D}}_{t}}$, respectively. Both corpora have the same predefined $C$ emotion classes. In source-free cross-corpus SER setting, the source corpus ${{\mathcal{D}}_{s}}$ is only available for source model pre-training, while in target adaptation we merely access the pre-trained source model and the unlabeled target corpus ${{\mathcal{D}}_{t}}$. The feature extractor takes a speech sample ${x_i}$ as input and produces a feature representation denoted as ${\bm{f}_{i}}=f({{x}_{i}})\in {{\mathbb{R}}^{d}}$, where $d$ is the dimension of the feature space. The output of the classifier is denoted as ${{p}_{i}}=\sigma ({\bm{f}_{i}})\in {{\mathbb{R}}^{C}}$, where $\sigma (.)$ is the softmax function. Fig.\ \ref{fig1} shows the proposed structure.

\subsection{Nearest Neighbor Contrastive Learning}
\label{ssec:subhead}
As mentioned before, even though the source classifier may not be suitable for the target domain, target speech samples tend to form distinct clusters in the feature space. It indicates that similar samples are expected to be close to each other. 
Thus, we propose a nearest neighbor contractive learning module to leverage the local information within the target data and enhance semantic consistency. This module focuses on the adaptation between samples and aims to reinforce the relationships among nearest neighbors. 

To retrieve nearest neighbors during training, we build  a feature memory bank $\mathcal{F}=[\bm{f}_{1}, \bm{f}_{2}, ..., \bm{f}_{N_t}]$ to store features of $N_t$ target samples. The cosine similarity is used for retrieving $k$-nearest neighbors: 
\begin{equation}
\mathcal{N}_{K}^{i}=\{{{\mathcal{F}}_{j}}|top\mbox{-}k(\cos ({\bm{f}_{i}},{{\mathcal{F}}_{j}})),\forall {{\mathcal{F}}_{j}}\in \mathcal{F}\}
\end{equation}
It is important to note that before every batch training iteration, we update the feature bank $\mathcal{F}$ by replacing the existing items with their corresponding counterparts from the current batch. Additionally, we set the number of nearest neighbors $k$ for each sample as 1. 

We treat the nearest neighbors as positive pairs and other samples as negative ones. Based on the InfoNCE loss \cite{oord2018representation}, we define the nearest neighbor contrastive learning loss as:
\begin{equation}
\begin{aligned}
 {{\mathcal{L}}_{ncl}}=-\frac{1}{{{N}_{t}}}\sum\limits_{i=1}^{{{N}_{t}}}{\sum\limits_{{\bm{k}^{+}}\in \mathcal{N}_{K}^{i}}{\log \frac{\exp (\phi ({\bm{f}_{i}},{\bm{k}^{+}})/\tau \ )}{\sum\limits_{{\bm{k}_{j}}\in \mathcal{N}_{f_i}}{\exp (\phi ({\bm{f}_{i}},{\bm{k}_{j}})/\tau \ )}}}},  
\end{aligned}
\label{eq1}
\end{equation}
where ${\bm{k}^{+}}$ is the feature in the $k$-nearest neighbors set $\mathcal{N}_{K}^{i}$ of ${\bm{f}_{i}}$, and $\mathcal{N}_{f_i}$ denotes the feature set except ${\bm{f}_{i}}$. $\phi (.)$ denotes the cosine similarity. $\tau $ is a temperature hyper-parameter and is empirically set as 0.05. By minimizing the loss function ${{\mathcal{L}}_{ncl}}$, our ECAN can push features towards their nearest neighbors and pull them away from dissimilar ones. 

\subsection{Supervised Contrastive Learning}
\label{ssec:subhead}
In the nearest neighbor contrast approach, there is a possibility of encountering noisy neighbors that belong to different emotion categories. This can lead to incorrect supervision. To address this issue, we propose a supervised contrastive learning module, which aims to bring features belonging to the same category closer and push features from different categories farther apart. To implement this, a score bank $\mathcal{S}=[\bm{p}_{1}, \bm{p}_{2}, ..., \bm{p}_{N_t}]$ is introduced to store the softmax prediction scores of all target data points. Similar to the feature bank $\mathcal{F}$, the score bank $\mathcal{S}$ is updated before each batch training iteration. In the supervised contrastive learning process, the samples with the same category are searched within the score bank $\mathcal{S}$. The corresponding features in the feature bank $\mathcal{F}$ are extracted to form the $i^{th}$ emotion feature set ${{\mathcal{C}}_{i}}$. We define the supervised contrastive learning loss ${{\mathcal{L}}_{scl}}$ as follows:
\begin{equation}
\begin{aligned}
{{\mathcal{L}}_{scl}}=\sum\limits_{i=1}^{{{N}_{t}}}\frac{-1}{N_{\mathcal{C}_{i}}}{\sum\limits_{{\bm{q}^{+}}\in {{\mathcal{C}}_{i}}}{\log \frac{\exp (\phi ({\bm{f}_{i}},{\bm{q}^{+}})/\tau \ )}{\sum\limits_{{\bm{q}_{j}}\in \mathcal{N}_{f_i}}{\exp (\phi ({\bm{f}_{i}},{\bm{q}_{j}})/\tau \ )}}}},  
\end{aligned}
\label{eq2}
\end{equation}
where ${\bm{q}^{+}}$ is the feature in the ${{\mathcal{C}}_{i}}$, and  $N_{\mathcal{C}_{i}}$ is the number of features in ${{\mathcal{C}}_{i}}$. Through optimizing ${{\mathcal{L}}_{scl}}$, ECAN reduces the impact of unreasonable nearest neighbors via enhancing the inter-class discrimination and intra-class compactness of target features.
\begin{algorithm}
	\renewcommand{\algorithmicrequire}{\textbf{Input:}}
	\renewcommand{\algorithmicensure}{\textbf{Output:}}
	\caption{Training procedure of the ECAN}
	\label{alg1}
	\begin{algorithmic}[1]
            \REQUIRE Source corpus ${{\mathcal{D}}_{s}}$ (only available during pre-training); Target corpus ${{\mathcal{D}}_{t}}$ 
		\STATE Pre-train source model on ${{\mathcal{D}}_{s}}$
		\STATE Build memory bank $\mathcal{F}$ and $\mathcal{S}$ for ${{\mathcal{D}}_{t}}$ based on the pre-trained model
		\WHILE{$t$ \textbf{in} $adaptation$-$iterations$}
		\STATE Sample a mini-batch $\mathcal{B}$ from ${{\mathcal{D}}_{t}}$
		\STATE Update $\mathcal{F}$ and $\mathcal{S}$ with current mini-batch $\mathcal{B}$
		\STATE Compute loss function ${{\mathcal{L}}_{ncl}}$ based on $\mathcal{F}$
		\STATE Compute loss function ${{\mathcal{L}}_{scl}}$ based on $\mathcal{F}$ and $\mathcal{S}$
        \STATE Compute loss function ${{\mathcal{L}}_{div}}$ and obtain the total loss $\mathcal{L}$
        \STATE Update the model based on SGD algorithm
        \STATE $t \leftarrow t + 1$
	  \ENDWHILE  
	\end{algorithmic}  
\end{algorithm}

\subsection{Total Objective}
\label{ssec:subhead}
We further encourage the diversity in model predictions to avoid collapsing solutions, where the model predicts some specific classes for all target samples. The diversity loss ${{\mathcal{L}}_{div}}$ that encourages the prediction balance is defined below:
\begin{equation}
\begin{aligned}
{{\mathcal{L}}_{div}}=\sum\limits_{c=1}^{C}{\mathrm{KL}({{{\bar{p}}}_{c}}||\frac{1}{C})},\text{ with }{{\bar{p}}_{c}}=\frac{1}{{{N}_{t}}}\sum\limits_{i=1}^{{{N}_{t}}}{p_{i}^{(c)}},  
\end{aligned}
\label{eq3}
\end{equation}
where $\mathrm{KL}\left( \cdot ||\cdot  \right)$ denotes the Kullback-Leibler divergence \cite{van2014renyi}. The $p_{i}^{(c)}$ is the predicted score of the $c^{th}$ emotion class of $x_{i}$, and ${{\bar{p}}_{c}}$ represents the predicted probability of class $c$, which is regularized by the uniform distribution.

Finally, the total adaptive loss function combining Eqs.\ (\ref{eq1}), (\ref{eq2}), and (\ref{eq3}) is as follows:
\begin{equation}
\begin{aligned}
\mathcal{L}={{\mathcal{L}}_{div}}+\lambda {{\mathcal{L}}_{ncl}}+\beta {{\mathcal{L}}_{scl}},
\end{aligned}
\label{eq4}
\end{equation}
where $\lambda $, $\beta $ are trade-off coefficients to balance local structural clustering and emotion discrimination. The training process of our proposed ECAN is illustrated in Algorithm 1.

\section{Experiments}
\label{sec:majhead}

\begin{table}[t!]
\centering
\renewcommand{\arraystretch}{1.3}
\caption{The sample statistics of corpora used in all cross-corpus SER tasks.}
\resizebox{0.48\textwidth}{!}{
\begin{tabular}{|c|c|c|c|c|c|c|c|c|}
\hline
\textbf{Corpus} & Anger & Sad & Fear & Happy & Disgust & Neutral & Surprise & Total \\ \hline
EMOVO& 84  & 84 & 84 & 84 & 84 & 84 & 84 & 588 \\ \hline
EmoDB & 127 & 62 & 69 & 71 & 46 & 79 & - & 454 \\ \hline
eNTERFACE& 215 & 215 & 215 & 215 & 212 & - & 215 & 1287 \\ \hline
CASIA& 200 & 200 & 200 & 200 & - & 200 & 200 & 1200 \\ \hline
\end{tabular}}
\label{tab1}
\end{table}

\begin{table*}[t!]
\centering
\small 
\renewcommand{\arraystretch}{1.1}
\caption{UARs of State-of-the-arts, where the best results are highlighted in bold.(\%)}
\resizebox{0.98\textwidth}{!}{
\begin{tabular}{c|c|cccccccccccc|c}
\hline
\textbf{Methods} & Source-free & \textbf{B$\rightarrow$C} & \textbf{C$\rightarrow$B} & \textbf{B$\rightarrow$E} & \textbf{E$\rightarrow$B} & \textbf{C$\rightarrow$E} & \textbf{E$\rightarrow$C} & \textbf{B$\rightarrow$O} & \textbf{O$\rightarrow$B} & \textbf{C$\rightarrow$O} & \textbf{O$\rightarrow$C} & \textbf{E$\rightarrow$O} & \textbf{O$\rightarrow$E} & \textbf{Avg.}\\
\hline 
Source Only &-& 30.10 & 41.78 & 25.34 & 27.83 & 20.75 & 22.90 & 30.24 & 20.83 & 25.79 & 25.92 & 25.20 & 22.35 & 26.59 \\
DAN (ICML' 15) &\ding{56}& 36.30 & 56.72 & 33.58 & 43.50 & 32.17 & 29.30 & 35.53 & 44.24 & \textbf{36.51} & \textbf{32.42} & \textbf{32.14} & 28.93 & 36.78 \\
\hline
 SHOT (ICML' 20) &\ding{52}& 34.80 & 55.83 & 31.92 & 46.08 & 30.99 & 31.80 & 35.52 & 44.63 & 34.13 & 30.00 & 27.98 & 28.76 & 36.04 \\
 G-SFDA (ICCV' 21) &\ding{52}& 27.90 & 50.41 & 23.21 & 35.79 & 27.09 & 24.30 & 25.20 & 36.32 & 24.40 & 20.42 & 25.00 & 25.36 & 28.78 \\
 NRC (NeurIPS' 21) &\ding{52}& 37.80  & 60.16 & 31.87 & 48.42 & 32.22 & 31.40 & 35.12 & \textbf{44.93} & 32.74 & 29.00 & 29.76 & 27.73 & 36.76 \\
 USFAN (ECCV' 22) &\ding{52}& 32.70 & 53.72 & 30.77 & \textbf{51.73} & 30.60 & 29.00 & 33.93 & 44.65 & 31.94 & 30.08 & 29.17 & 27.53 & 35.49 \\
 CoWA-JMDS (ICML' 22) &\ding{52}& 34.50 & 55.26 & 28.49 & 36.74 & 31.61 & 25.80 & 29.96 & 27.78 & 29.96 & 25.75 & 25.99 & 24.08 & 31.33 \\
 DaC (NeurIPS' 22)  &\ding{52}& 31.90 & 47.42 & 26.42 & 39.03 & 25.60 & 27.00 & 24.40 & 30.19 & 25.00 & 25.83 & 28.37 & 24.63 & 29.65 \\
 AaD (NeurIPS' 22)  &\ding{52}& 35.00 & 55.50 & 31.47 & 48.12 & 32.17 & 29.10 & 34.52 & \textbf{44.93} & 34.92 & 27.30 & 25.60 & 26.50 & 35.43\\
\hline
ECAN (Ours)  &\ding{52}& \textbf{39.00} & \textbf{61.37} & \textbf{34.21} & 46.87 & \textbf{34.53} & \textbf{31.90} & \textbf{36.51} & 40.86 & 35.91 & 27.42 & 28.57 & \textbf{29.16} & \textbf{37.19} \\
\hline
\end{tabular}}
\label{tab2}
\end{table*}

\subsection{Corpora and Evaluation Protocols}

We utilize four public available speech emotion corpora, \emph{i.e.}, EMOVO (O) \cite{costantini2014emovo}, EmoDB (B)~\cite{2005A}, eNTERFACE (E)~\cite{DBLP:conf/icde/MartinKMP06}, and CASIA (C)~\cite{zhang2008design}, for evaluations. EMOVO is an emotional corpus applicable to Italian. It is recorded by 6 actors, containing 7 emotions (\textit{anger}, \textit{sad}, \textit{fear}, \textit{happy}, \textit{disgust}, \textit{neutral} and \textit{surprise}). EmoDB is a German speech emotion corpus. It consists of 535 speech samples and 10 speakers are required to perform 7 emotional scripts (\textit{angry}, \textit{disgust}, \textit{sad}, \textit{happy}, \textit{fear}, \textit{neutral}, and \textit{bored}). As for eNTERFACE, which is an English audio-visual emotion database, It is collected from 43 individuals including 1582 samples. Samples are labeled as one of 6 basic emotions including \textit{happy}, \textit{sad}, \textit{fear}, \textit{angry}, \textit{surprise} and \textit{disgust}. We adopt the audio data in the experiments. CASIA is a Chinese speech emotion corpus which is built from 4  speakers with 1200 audio samples. Each speaker is required to perform utterances under 6 different emotion states, \emph{i.e.}, \textit{anger}, \textit{sad}, \textit{fear}, \textit{happy}, \textit{neutral} and \textit{surprise}. 

In source-free cross-corpus SER, one speech corpus serves as the source one and the other corpus is the target one. By alternatively choosing either two of the above four speech emotion corpora, we design twelve tasks. The sample statistics is shown in Tab \ref{tab1}. For evaluation metric, we adopt unweighted average recall (UAR), which is defined as the average of the prediction accuracy per class. We also report the average results of all tasks.

\subsection{Baselines and Implementation Details}
To validate the effectiveness of our proposed ECAN, we adopt several domain adaptation methods as baselines for comparison, \emph{i.e.}, SHOT \cite{liang2020we}, NRC \cite{yang2021exploiting}, DAN \cite{long2015learning}, G-SFDA \cite{YangW0HJ21}, CoWA-JMDS \cite{0004JYY22}, USFAN \cite{RoyTPKSRS22}, DaC \cite{ZhangCCLLLL22}, and AaD \cite{YangWWJ022}. Besides, we also compare to the source-only model, which refers to the pre-trained model by source data.

For the aforementioned methods, we refer to the official code implementation to conduct experiments. To ensure a fair comparison, we choose VGG-11 \cite{SimonyanZ14a} as the backbone architecture and leverage the Mel spectrum with a size of 224 $\times$ 224 as the network input. For pre-training the source model, we employ label smoothing and stochastic gradient descent with a momentum of 0.9. The training process is conducted for 100 epochs. To obtain the optimal results, we perform a parameter search by exploring different trade-offs within fixed intervals for the mentioned methods. Specifically, for SHOT, USFAN and DAN, the search interval is \{0.0001:0.0001:0.001, 0.001:0.001:0.01, 0.01:0.01:0.1, 0.1:0.1:1, 2, 5, 10, 100\}. For G-SFDA and AaD, we search the nearest neighbor number from [1:1:10]. For NRC, both numbers of reciprocal neighbor and expanded neighbor are searched from [1:2:9]. The mixup weight parameter set of CoWA-JMDS is \{0.1:0.1:1, 5, 50, 100\}. As for ECAN and DaC, $\lambda$ and $\beta$ are searched from \{0.00001, 0.00005, 0.0001, 0.0005, 0.001, 0.005, 0.01, 0.05, 0.1, 0.5, 1, 5, 10, 50, 100\} and \{0.1, 0.3, 0.6, 0.9, 2\}, respectively.

\subsection{Comparison with State-of-the-arts}
Experimental results of different methods are reported in Tab \ref{tab2}. We can observe some interesting findings. Firstly, it can be obviously seen that our proposed ECAN achieves the best performance compared to all other comparison methods, where the average UAR reaches 37.19\%. Diving deeper into each task, our ECAN outperforms other methods in seven out of twelve cross-corpus SER tasks, especially in B$\rightarrow$C (39.00\%) and C$\rightarrow$E (34.53\%) tasks. Despite not achieving the best results in the remaining tasks, ECAN is still competitive with the best performing methods, \emph{e.g.}, in the C$\rightarrow$O task (ECAN 35.91\% \emph{v.s.} DAN 36.51\%). Moreover, ECAN remains superior even when compared to the DAN method that has access to source corpus data. These findings demonstrate the effectiveness of our proposed ECAN in dealing with the source-free cross-corpus SER problem.


Based on the experimental results, it is evident that most methods struggle to perform well on the C$\rightarrow$E and E$\rightarrow$C tasks. It can be attributed to the variations in emotion induction methods and recorded language between the  CASIA and eNTERFACE datasets. Specifically, CASIA consists of Chinese corpora where the speech samples are acted by the speakers, while eNTERFACE comprises English corpora where emotions are induced using pre-prepared materials. A similar trend can also be observed in the tasks between EMOVO and eNTERFACE (E$\rightarrow$O and O$\rightarrow$E tasks), where EMOVO consists of acted Italian that significantly differs from eNTERFACE. These differences in language, emotion induction techniques, and cultural nuances could affect the performance of the methods on these specific tasks.

\subsection{Comparison with Cross-Corpus SER Methods}
\begin{figure}[t!] 
\centering 
\includegraphics[width=0.28\textwidth]{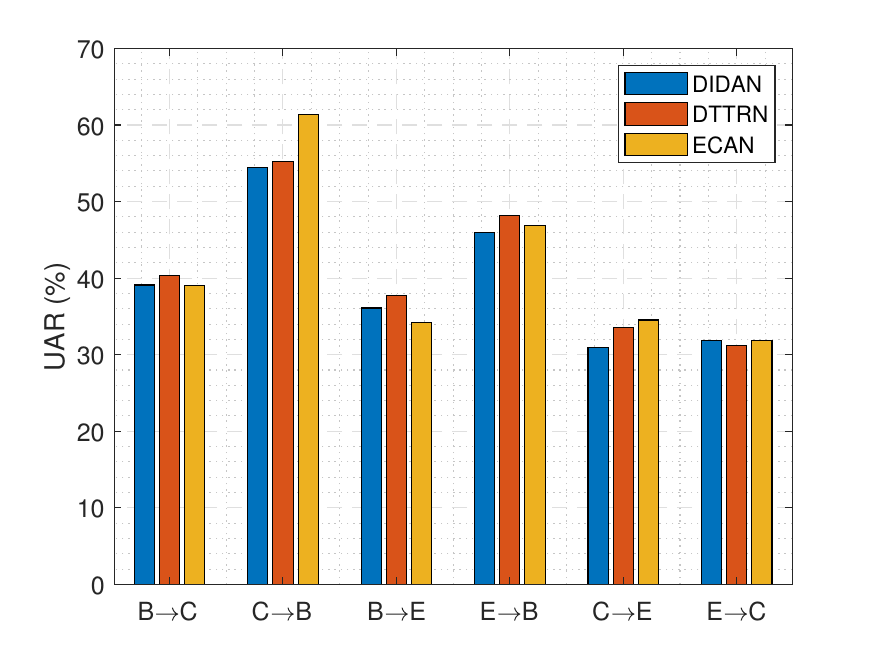}\\ 
\caption{Comparison with Cross-Corpus SER Methods.}
\label{fig2} 
\end{figure}

In this section, we conduct a comparison between our proposed ECAN and previous cross-corpus SER algorithms on six tasks, as depicted in Fig.\  \ref{fig2}. Notably, even without the presence of source data, ECAN achieves comparable performance to source-available methods such as DIDAN \cite{10095388} and DTTRN \cite{zhao2022deep}. Specifically, our ECAN surpasses methods that utilize source data in the C$\rightarrow$B and C$\rightarrow$E tasks, while slightly trailing behind in the remaining tasks. These results highlight the superiority of our proposed ECAN algorithm in successfully addressing cross-corpus SER challenges.


\subsection{Ablation Study}
To evaluate the effectiveness of our proposed functions, we perform ablation experiments on six tasks. Tab \ref{tab3} shows results of the ablation study. It is evident that when ${{\mathsf{\mathcal{L}}}_{ncl}}$ and ${{\mathsf{\mathcal{L}}}_{scl}}$ are considered together, the performance is consistently better compared to using either loss alone. Moreover, the results decrease when ${{\mathsf{\mathcal{L}}}_{div}}$ is removed, indicating the importance of ${{\mathsf{\mathcal{L}}}_{div}}$ that promotes diversity among samples.
\begin{table}[t!]
\footnotesize
\centering
\renewcommand{\arraystretch}{1.1}
\caption{The ablation study for the proposed ECAN.(\%)}
\resizebox{0.48\textwidth}{!}{
\begin{tabular}{|c|cccccc|}
\hline
\textbf{Methods} & \textbf{B$\rightarrow$C} & \textbf{E$\rightarrow$B} & \textbf{C$\rightarrow$E} & \textbf{B$\rightarrow$O} & \textbf{C$\rightarrow$O} & \textbf{O$\rightarrow$E}   \\ \hline
ECAN \emph{w/o} $\mathcal{L}_{ncl}$ & 37.20  & 41.32 & 32.26 & 32.54 & 30.36 & 26.38 \\ 
ECAN \emph{w/o} $\mathcal{L}_{scl}$  & 33.00 & 42.27 & 30.00 & 34.92 & 30.75 & 25.67 \\ 
ECAN \emph{w/o} $\mathcal{L}_{div}$ & 36.80 & 45.54 & 32.08 & 35.12 & 33.73 & 28.85  \\ 
ECAN & \textbf{39.00} & \textbf{46.87} & \textbf{34.53} & \textbf{36.51} & \textbf{35.91} & \textbf{29.16}  \\ \hline
\end{tabular}}
\label{tab3}
\end{table}


\subsection{Feature Visualization}

\begin{figure}[t!] 
\centering 
\includegraphics[width=0.4\textwidth]{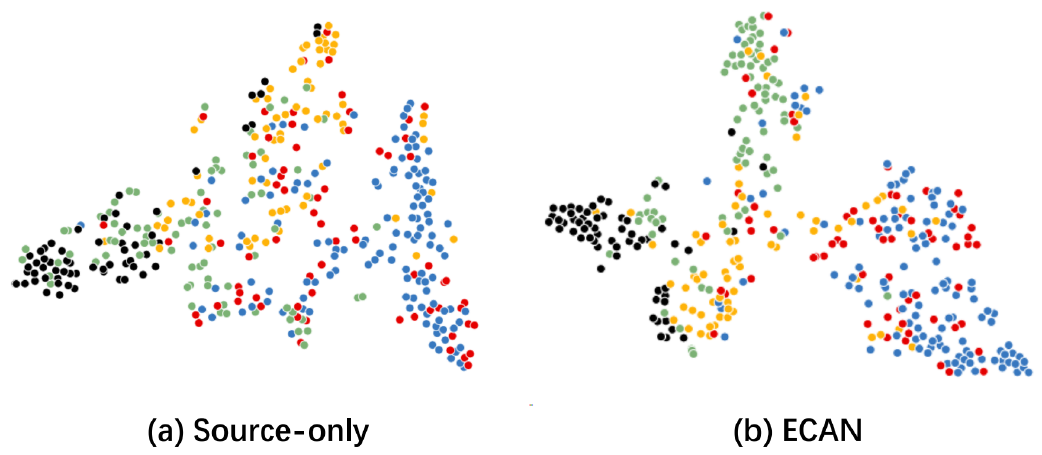}\\ 
\caption{The t-SNE visualization on the task of C$\rightarrow$B.}
\label{fig3} 
\end{figure}

To provide a more visually compelling demonstration of our algorithm's performance, we employ t-SNE \cite{van2008visualizing} for feature visualization. Specifically, we focus on the features extracted by our model in the C$\rightarrow$B task and present the results in Fig.\  \ref{fig3}. Each colored dot represents a different emotion category. From Fig.\ \ref{fig3}(a), we can observe that the target features are initially scattered in the feature space, which can be attributed to the domain shift before adaptation. However, after the adaptation, the target features form clearer and more compact clusters that correspond to specific emotion categories. This outcome clearly demonstrates the effectiveness of our proposed strategy, which simultaneously considers nearest neighbors and separates clusters of different emotion categories. The improved clustering of target features showcases the efficacy of our algorithm in addressing the challenges of cross-corpus SER tasks.

\section{Conclusion}
In this paper, we proposed a simple yet effective method called ECAN to solve a practical and previously unexplored source-free cross-corpus SER problem. ECAN leverages the local structure of the target data by utilizing nearest neighbor contrastive learning. This approach allows for successful adaptation without relying on the source data. Besides, ECAN enhances class-level adaptation by aggregating features within the same emotion category and separating different emotion clusters to create clear classification boundaries. Extensive experiments were conducted on four widely used speech emotion corpora, and the results verify the effectiveness of our proposed ECAN in dealing with the source-free cross-corpus SER task. In future work, we will further explore neighboring information and compare with other advanced algorithms. 

\vfill\pagebreak

%

\small
\bibliographystyle{IEEEbib}
\bibliography{refs}

\end{document}